\documentclass[aps,prd,nofootinbib]{revtex4}
\usepackage{graphicx}
\begin{document}
\title{The clustering of ultra-high energy cosmic rays and their sources} 
\author{N.W. Evans $^1$, F. Ferrer $^2$
and S. Sarkar $^2$}
\medskip
\affiliation{$^1$ Institute of Astronomy, University of Cambridge,
Madingley Road, Cambridge CB3 OHA, UK \\ 
$^2$ Theoretical Physics, University of Oxford, 1 Keble Road,
Oxford OX1 3NP, UK} 
%\date{\today}

\begin{abstract}
The sky distribution of cosmic rays with energies above the `GZK
cutoff' holds important clues to their origin. The AGASA data,
although consistent with isotropy overall, shows evidence for
small-angle clustering, and it has been argued that such clusters are
aligned with BL Lacertae objects, implicating these as the sources. It
has also been suggested that such clusters can arise if the cosmic
rays come from the decays of very massive relic particles in the
Galactic halo, due to the expected clumping of cold dark matter. We
examine these claims and show that both are, in fact, unjustified.
\end{abstract}

\maketitle

\section{Introduction}
The mystery of the ultra-high energy cosmic rays (UHECRs) with
energies exceeding $E_{\rm GZK}\simeq4\times10^{19}$~eV --- the
Greisen-Zatsepin-Kuzmin (GZK) `cutoff'
\cite{Greisen:1966jv,Zatsepin:1966jv} --- continues to deepen. This
energy sets the threshold for photomeson production on the cosmic
microwave background so the observation of such UHECRs (assumed to be
protons or heavier nuclei) would indicate that the sources are
relatively nearby, within the local supercluster of galaxies
\cite{Stecker:1968uc}. Recent observations by the HiRes air
fluorescence detector \cite{Abu-Zayyad:2002sf} are however
inconsistent with previously published data from the Akeno Giant Air
Shower Array (AGASA) which ruled out such a cutoff with a significance
$\gtrsim5\sigma$ \cite{Takeda:1998ps,Hayashida:2000zr}. HiRes has
reported only 1 event above $10^{20}$~eV, whereas about 20 would have
been expected on the basis of the AGASA spectrum. The two spectra can
be made to agree {\em below} this energy, if the energies of the AGASA
events are systematically lowered by 20\% (within the quoted
uncertainty), however 5 of them still remain above this energy
\cite{Bergman:2002pw}. Subsequently the AGASA collaboration have
carefully assessed their energy measurement uncertainties and
reaffirmed that their observed spectrum does extend well beyond the
GZK energy \cite{Takeda:2002at}. To resolve this situation requires
making simultaneous measurements using both the air shower and air
fluorescence methods; such measurements are underway at the Pierre
Auger Observatory being constructed in Argentina
\cite{Watson:2001et,pao}.

Another development has been the AGASA observation that the UHECR
arrival directions, although consistent with isotropy overall, exhibit
clustering on small angular scales \cite{Takeda:1999sg}. Among the 59
AGASA events above $4\times10^{19}$~eV, there are 5 `doublets' and 1
`triplet' with separation angle less than the estimated angular
resolution of $2.5^\circ$ \cite{takeda}.\footnote{For simulated events
with $E>4\times10^{19}$~eV, 68\% have a reconstructed arrival
direction within $1.8^\circ$ of the true direction and 90\% within
$3^\circ$; the corresponding angles for all events above $10^{19}$~eV
are $2.8^\circ$ and $4.6^\circ$, keeping in mind that the energy
resolution is $\pm30\%$ \cite{Hayashida:2000zr,Takeda:2002at}.}  The
probability for this to arise by chance from an isotropic distribution
is less than 0.1\%. However this probability is very sensitive to the
assumed angular resolution \cite{Goldberg:2000zq}, e.g. increasing to
$\sim1\%$ if the angular resolution is $3^\circ$
\cite{Anchordoqui:2001qk}. Moreover adding data from three other air
shower experiments (Volcano Ranch, Haverah Park, and Yakutsk) {\em
dilutes} the significance. In an earlier such analysis
\cite{Uchihori:1999gu}, 8 doublets and 2 triplets were found in a
dataset of 47 AGASA plus 45 other events with $E>4\times10^{19}$~eV,
taking the effective angular resolution of the dataset to be
$4^\circ$. The chance probability for this to arise from an uniform
distribution is $\sim10\%$, thus statistically not significant.

Nevertheless, the existence of such clusters has been linked to the
possibility of (repeating) point sources of UHECR
\cite{Dubovsky:2000gv,Tinyakov:2001ic}, specifically cosmologically
distant BL Lacertae \cite{Tinyakov:2001nr} --- a sub-class of active
galactic nuclei (AGN) which have been long discussed as possible
accelerators of UHECRs \cite{Biermann:ep}. However the expected
deflections of UHECRs (assumed to be charged particles) by Galactic
and intergalactic magnetic fields ought to smear out such tight source
correlations
\cite{Stanev:1997,MedinaTanco:1997rt,Alvarez-Muniz:vf}.\footnote{In
fact focussing effects by such fields can give rise to apparent
clustering even when the arrival directions are random
\cite{MedinaTanco:1998yq,Harari:2000az,Harari:2002dy}.} Contrary to
these results, it has been claimed recently that the correlations with
BL Lacs are preserved, even improved, if the UHECRS are protons, after
allowing for deflections by the Galactic magnetic field
\cite{Tinyakov:2001ir}. Little is known about the intergalactic
magnetic field \cite{Widrow:2002ud}; requiring rectilinear propagation
of protons over the attenuation length of $l\sim1000$~Mpc at
$E>4\times10^{19}$~eV (decreasing to $\sim100$~Mpc at $E>10^{20}$~eV
\cite{Stanev:2000fb}) would imply that its homogeneous component on
such scales is extremely weak: $B<2\times10^{-12}(l/1000\,{\rm
Mpc})^{-1}$~G \cite{Berezinsky:2002vt}. It has also been claimed
\cite{Blasi:2000ud,Blasi:2001kr} that such clustering is predicted in
a model where the UHECR arise from the decay of superheavy relic
particles accumulated in the Galactic halo
\cite{Berezinsky:1997hy,Birkel:1998nx}, due to the expected clumping
of halo dark matter.

In this paper we examine both these claims in detail, using as our
basic statistical tool the two-point correlation function. Our
intention is to determine whether the claimed correlations are
meaningful, given the present limited event statistics.

\section{UHECR clustering and correlations with possible sources}

It is natural to look for correlations between the observed UHECR
arrival directions and plausible astrophysical sources, however it is
essential to take care not to generate spurious correlations by
introducing biases. For example it has been claimed that the 5 highest
energy events with $E>10^{20}$~eV are all aligned with compact
radio-loud quasars (CRQSOs) having redshifts between 0.3 and 2.2, and
the chance probability for this coincidence was estimated to be 0.5\%
\cite{Farrar:1998we,Farrar:1999fw}. However this rises to 3\% when the
event used to formulate the hypothesis itself (the previously noted
\cite{Elbert:1994zv} alignment of the quasar 3C147 with the
$3.2\times10^{20}$~eV Fly's Eye event \cite{Bird:1994uy}) is excluded
from the sample \cite{Hoffman:1999ev}. A careful recent analysis
\cite{Sigl:2000sn} based on an updated event list (5 AGASA
\cite{Hayashida:2000zr}, 4 Haverah Park \cite{Lawrence:cc,Ave:2000nd}
and 1 Fly's Eye \cite{Bird:1994uy}) demonstrates that there are {\em
no} significant correlations between UHECRs and CRQSOs. These authors
show also that another recent claim \cite{Virmani:xk} of significant
correlations with CRQSOs is based on inadequate data, and, in
addition, that there are {\em no} significant correlations with an
interesting sub-group of these sources, viz. $\gamma$-ray blazars
\cite{Sigl:2000sn}. A correlation between events with
$E>4\times10^{19}$~eV and nearby galaxies likely to host quasar
remnants (QRs) has also been found at the $3\sigma$ level, although
this disappears if attention is restricted to events above
$10^{20}$~eV \cite{Torres:2002bb}.

What has revived interest in the possibility of such correlations is
the claimed clustering in the arrival directions of UHECRs
\cite{Takeda:1999sg}. This may arise for example if the sources are
`compact' (i.e. smaller than the experimental angular resolution) with
the clusters corresponding to more than one UHECR being received from
the {\em same} source. Since the number of events in such clusters is
much smaller than the total number of events, the majority of such
sources have clearly not been seen at all. However it is possible to
estimate their number density using Poisson statistics. Allowing for
the attenuation of UHECRs from distant sources due to GZK energy
losses, the observed occurences of `triplets' and `doublets' relative
to the number of single events was used to estimate the spatial
density of such sources to be $6\times10^{-3}$ Mpc$^{-3}$
\cite{Dubovsky:2000gv}. This would obviously place stringent
constraints on candidate astrophysical sources, e.g. $\gamma$-ray
bursts (GRBs) have a spatial density of only $\sim10^{-5}$
Mpc$^{-3}$. However a more careful analysis \cite{Fodor:2000yi} shows
that the uncertainties in this estimate are very large. The true
number is $2.77_{-2.53(2.70)}^{+96.1(916)}\times10^{-3}$ Mpc$^{-3}$ at
the 68\% (95\%) c.l.; moreover relaxing the assumptions made,
viz. that the sources all have the same luminosity and a spectrum
$\propto\,E^{-2}$, increases the allowed range even further, e.g. to
$180_{-165(174)}^{+2730(8817)}\times10^{-3}$ Mpc$^{-3}$ for a
Schechter luminosity function and a spectrum $\propto\,E^{-3}$
\cite{Fodor:2000yi}. Clearly the present limited event numbers do {\em
not} permit any candidate class of sources to be definitively
excluded. Note that the observed clustering may also arise because of
a higher density of local sources in certain directions, e.g. due to
the clumpiness of halo dark matter in the decaying dark matter model.

The next step taken was the construction of the angular
autocorrelation function of UHECRs \cite{Tinyakov:2001ic}. For the
AGASA data this displays a clear peak at separation angles less than
$2.5^\circ$, consistent with the point spread function
\cite{takeda}. Moreover the chance probability estimated by Monte
Carlo to match or exceed the observed count in the first angular bin,
when plotted versus energy, is seen to have a minimum at $E_{\rm
GZK}$; the peak in the autocorrelation function for
$E>4\times10^{19}$~eV is stated to have a significance of 4.6$\sigma$
\cite{takeda}. An equally significant autocorrelation was claimed
using a different method of analysing the data in which a `triplet'
was taken to correspond to three or two `doublets' depending on
whether the events are bunched together or linearly aligned
\cite{Tinyakov:2001ic}. These authors found the probability for chance
coincidences to be minimum for events above $4.8\times10^{19}$~eV in
the AGASA data \cite{Hayashida:2000zr}, and above
$2.4\times10^{19}$~eV in the Yakutsk data \cite{yakutsk}. Restricting
attention to events above these energies, the chance probability for
the observed clustering in the first angular bin was quoted as
$3\times10^{-4}$ for AGASA and $2\times10^{-3}$ for Yakutsk, taking
$2.5^\circ$ and $4^\circ$ respectively for the size of the first bin,
corresponding to the respective experimental angular resolutions
\cite{Tinyakov:2001ic}.

\subsection{Correlation with BL Lacs}

Motivated by the results quoted above which implicate compact sources
for UHECRs, Tinyakov \& Tkachev \cite{Tinyakov:2001nr} have proposed
that the sources are in fact BL Lacs. The physical motivation they
provided for this is that only AGNs in which the central jet points
towards us --- `blazars' --- are likely to be UHECR sources (since
particles accelerated in a relativistic jet are strongly beamed), and
among all blazars, BL Lacs in particular have few emission lines in
their spectra, indicating low density of ambient matter and radiation,
thus presumably more favorable conditions for particle acceleration.

Tinyakov \& Tkachev \cite{Tinyakov:2001nr} used a catalogue of AGNs
and quasars containing 306 confirmed (out of 462 listed) BL Lacs
\cite{veron2000}.
%\footnote{The Introduction to this states: {\em ``This catalogue
%should not be used for any statistical analysis as it is not complete
%in any sense, except that it is, hopefully, a complete survey of the
%literature''} \cite{veron2000}.}
They asserted that since the ability of BL Lacs to
accelerate UHECRs may be correlated with optical and radio emissions,
it would be appropriate to select the most {\em powerful} BL Lacs by
imposing cuts on redshift, apparent magnitude and 6~cm radio flux. In
fact the redshift is unknown for over half of all confirmed BL Lacs
but they assumed that all such BL Lacs are at $z>0.2$ and included
them in the sample anyway. By imposing the cuts $z>0.1$ (or unknown),
$m<18$, and $F_6>0.17$~Jy, they selected a sample of 22 BL Lacs.

They considered 39 AGASA events with $E>4.8\times10^{19}$~eV and 26
Yakutsk events with $E>2.4\times10^{19}$~eV, the energy cuts being
motivated by their earlier autocorrelation analysis
\cite{Tinyakov:2001ic} which had indicated that the small-angle
clustering of UHECRs is most pronounced above these energies in the
respective datasets. Assuming that the event energies are not
important for small angle correlations, they combined these into one
set of 65 events. Then they computed the correlation between the
arrival directions of these UHECRs and the selected 22 BL Lacs,
finding a significant number of coincidences. Eight UHECRs were found
to be within $2.5^\circ$ of 5 BL Lacs, the chance probability of which
is only $2\times10^{-5}$.\footnote{The most significant correlation
listed is the alignment of a BL Lac (1ES 0806+524) with a `triplet' in
the Yakutsk data having energies of (3.4, 2.8, 2.5)
$\times10^{19}$~eV. Note that these events are all {\em below} the GZK
cutoff. Moreover the uncertainty in the arrival direction of Yakutsk
events is $4^\circ$ at $4\times10^{19}$~eV, and even higher at lower
energies, so these close alignments are unlikely to be physically
significant.} The authors acknowledged that the imposition of the
arbitrary cuts made on the BL Lac catalogue can affect the
significance of this result and estimated the `penalty factor' to be
about 15; however the significance of the coincidences taking this
into account was then quoted as $6\times10^{-5}$ (implying a penalty
factor of only 3) \cite{Tinyakov:2001nr}. This was the basis for their
claim that BL Lacs are the probable sources of UHECRs.

Since these are cosmologically distant sources, it is pertinent to ask
how the UHECRs get to the Earth. Initially Tinyakov \& Tkachev
\cite{Tinyakov:2001nr} inferred that the primaries have to be neutral,
i.e. photons or neutrinos (unless the GZK effect is inoperative
because of violation of Lorentz invariance). However in subsequent
work \cite{Tinyakov:2001ir} they found that the correlations are {\em
improved} if the primaries are assumed to be protons, whose
trajectories are modified by the Galactic magnetic field (GMF). In
this work they used the full set of 57 AGASA events with
$E>4\times10^{19}$~eV (but no Yakutsk events) and allowed for
deflections by the regular component of the GMF (but ignored the
fluctuating component which is in fact of comparable strength
\cite{Beck:2000dc}), while assuming that deflections by intergalactic
magnetic fields (IGMF) are {\em negligible}. The same BL Lac catalogue
\cite{veron2000} was used but this time no cuts were made on redshift
or the 6~cm radio flux, only on the apparent magnitude ($m<18$) since
this maximised the correlations. It was found that 18 BL Lacs then lie
within $2.5^\circ$ of the reconstructed arrival directions of 22
UHECRs, if these mainly have charge +1 (however 8 might alternatively
be neutral and 4 must be neutral). Of these 18 BL Lacs, only 6 have
measured redshifts and these authors now proposed
\cite{Tinyakov:2001ir}, contrary to their previous supposition, that
the rest must in fact have redshifts {\em less than} 0.1 in order that
the protons they emit can overcome the GZK losses and reach the
Earth.\footnote{However of the 6 BL Lacs at known distances, only
B2~0804+35 ($z=0.082$) is near enough to be a possible source of the
$4.09\times10^{19}$ eV event (assumed to be charge +1) it is
associated with; the other object (TXS~0806+315, $z=0.22$) which is
also closely aligned with this event is probably too far on the basis
of UHECR propagation calculations \cite{Stanev:2000fb}. The remaining
4 pairings are also implausible --- these are RX~J1058.6+5628
($z=0.144$) aligned with a `doublet' (charge 0) having energies
$(5.35, 7.76)\times10^{19}$ eV, TEX~1428+370 ($z=0.564$) aligned with
an event (charge +1) of energy $4.97\times10^{19}$ eV, 1ES~1853+671
($z=0.212$) aligned with an event (charge +1) of energy
$4.39\times10^{19}$ eV, and EXO~1118.0+4228 ($z=0.124$) aligned with
an event (charge 0 or +1) of energy $7.21\times10^{19}$ eV. We have
estimated distances from the redshifts using the Mattig formula,
$d\sim4500\,{\rm Mpc}[z-0.2z^2]$, indicated by measurements of
cosmological parameters.} They asserted further \cite{Tinyakov:2001ir}
that their success at finding significant correlations between BL Lacs
and UHECRs in this manner confirmed that BL Lacs are the sources, {\em
as well} as validating their adopted model of the GMF, {\em and} their
assumption that there are no significant deflections due to the IGMF.
%\footnote{{\em ``There is something fascinating about
%science. One gets such wholesale returns of conjecture out of such a
%trifling investment of fact''} \cite{twain}.}

Moreover Tinyakov \& Tkachev \cite{Tinyakov:2001ir} noted that many of
these 22 BL Lacs are X-ray sources. In subsequent work
\cite{Gorbunov:2002hk}, an updated catalogue of QSOs containing 350
confirmed BL Lacs \cite{veron2001} was examined for correlations with
the third EGRET catalogue of $\gamma$-ray sources \cite{egret} and 14
were identified as strong $\gamma$-ray emitters (of which 8 were known
to be so already). Correlations between these 14 BL Lacs and the set
of 39 AGASA plus 26 Yakutsk events selected earlier
\cite{Tinyakov:2001nr} were then studied, again allowing for
deflections by the GMF modelled as in earlier work
\cite{Tinyakov:2001ir}. It was found that there are 13 possible
coincidences within $2.7^\circ$ (for charge 0 or +1) with a chance
probability of $3\times10^{-7}$ \cite{Gorbunov:2002hk}. Leaving out
the 2 BL Lacs that are invisible to the Northern Hemisphere cosmic ray
experiments, 8 of the remaining 12 are found to be along the
(reconstructed) arrival directions of UHECRs. It was concluded
\cite{Gorbunov:2002hk} that $\gamma$-ray emission is the physical
criterion for a BL Lac to be a UHECR source. However UHECRs are known
{\em not} to correlate with $\gamma$-ray blazars \cite{Sigl:2000sn}.
It was stated that there is no contradiction because the BL Lacs
considered display a low degree of polarisation whereas $\gamma$-ray
blazars are highly polarised \cite{Gorbunov:2002hk}.

Given this set of interesting claims we wish to ascertain to what
extent the strong correlations found depend on the selection criteria
used. To do so we calculate the correlation function in the same
manner as Tinyakov \& Tkachev
\cite{Tinyakov:2001ic,Tinyakov:2001nr,Tinyakov:2001ir} and use Monte
Carlo simulations to determine the probability of chance
coincidences. We consider four cases using the AGASA data
\cite{Hayashida:2000zr}; we do not consider the Yakutsk data
\cite{yakutsk} because the events which contribute dominantly to the
correlations found earlier
\cite{Tinyakov:2001ic,Tinyakov:2001nr,Gorbunov:2002hk} have energies
{\em below} the GZK cutoff, where the uncertainty in the arrival
directions exceeds $4^\circ$, so the correlations found at smaller
angles cannot be meaningful.

Our 4 models correspond to considering:
\begin{trivlist}

\item (1) the 39 UHECR with $E>4.8\times10^{19}$~eV
\cite{Hayashida:2000zr} and 22 BL Lacs selected by the Tinyakov \&
Tkachev criteria \cite{Tinyakov:2001nr},

\item (2) the full set of 57 UHECR with $E>4\times10^{19}$~eV
\cite{Hayashida:2000zr}, but retaining the cuts on the BL Lacs
\cite{Tinyakov:2001nr},

\item (3) the full set of 57 UHECR with $E>4\times10^{19}$~eV
\cite{Hayashida:2000zr} and the full set of 306 BL Lacs with no cuts
\cite{veron2000},

\item (4) the full set of 57 UHECR with $E>4\times10^{19}$~eV
\cite{Hayashida:2000zr} and 915 GRBs \cite{grb}.

\end{trivlist}
The last case is a control to determine whether there are equally
significant correlations with other suggested sources of UHECRs at
cosmological distances, which are {\em not} expected to contribute
events beyond the GZK cutoff \cite{Bahcall:2002wi}.

\begin{figure}[hbt]
\begin{center}
\begin{tabular}{c}
\begin{tabular}{c@{\hspace{2cm}}c}
(a)\includegraphics[width=7.5cm,height=4cm]{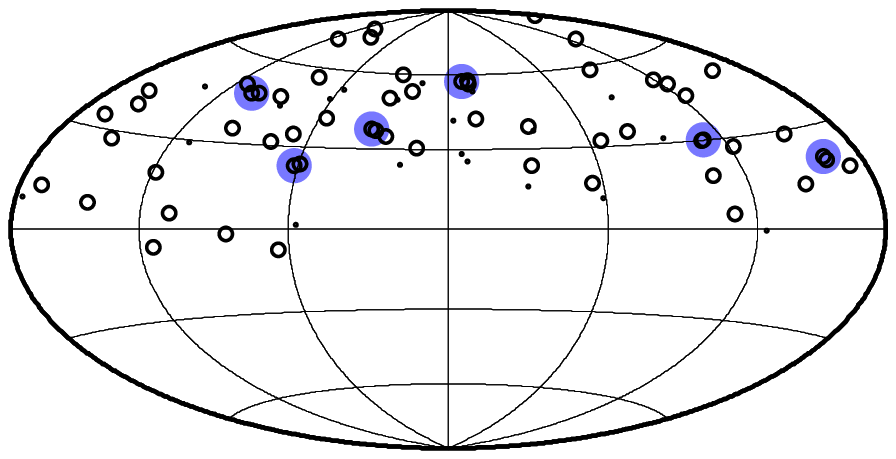}&
(b)\includegraphics[width=7.5cm,height=4cm]{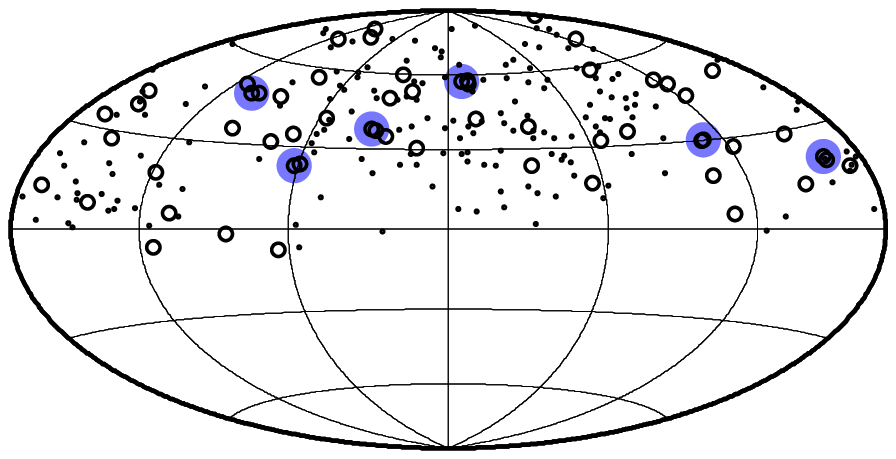}
\end{tabular} \\ \\
(c)\includegraphics[width=7.5cm,height=4cm]{fig1c.ps}
\end{tabular}
\end{center}
\caption{The sky distribution of 57 UHECRs (circles) with
$E>4\times10^{19}$~eV observed by AGASA
\protect\cite{Hayashida:2000zr} is shown, with the 5 `doublets' and 1
`triplet' marked with blue circles. Panel (a) shows also the 22 BL
Lacs (dots) satisfying the cuts on redshift, magnitude and 6~cm radio
flux imposed by Tinyakov \& Tkachev \protect\cite{Tinyakov:2001nr},
while panel (b) shows all 306 BL Lacs in the catalogue
\protect\cite{veron2001}. Panel (c) shows instead the 915 GRBs
observed by BATSE \protect\cite{grb}.}
\label{fig:skybl}
\end{figure}

\begin{figure}[hbt]
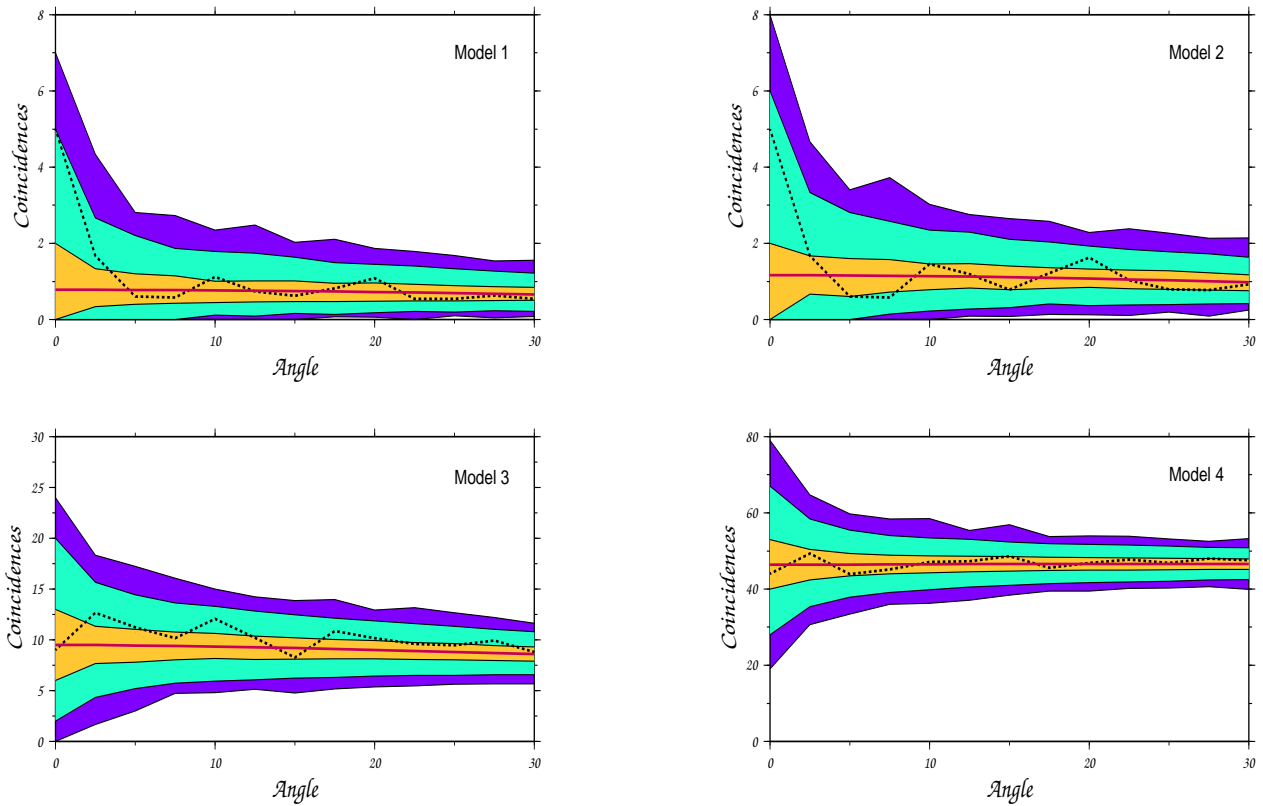

\begin{center}
\begin{tabular}{c@{\hspace{2cm}}c}
\includegraphics[width=7.5cm,height=5.5cm]{fig2a.ps}&
\includegraphics[width=7.5cm,height=5.5cm]{fig2b.ps}\\
\includegraphics[width=7.5cm,height=5.5cm]{fig2c.ps}&
\includegraphics[width=7.5cm,height=5.5cm]{fig2d.ps}
\end{tabular}
\end{center}
\caption{ The two-point correlation function between UHECRs observed
by AGASA and BL Lacs for the 4 models discussed, viz. using (1) the
energy cuts and selection criteria employed by Tinyakov \& Tkachev
\protect\cite{Tinyakov:2001nr}, (2) relaxing the cut in energy but
retaining the selection criteria for BL Lacs, (3) with no cuts at all,
and (4) using a catalogue of GRBs instead of BL Lacs. The regions
shaded light, medium and dark correspond to chance probabilities less
than 31.73\%, 0.27\% and 0.0057\% (chosen so as to correspond to
$1\sigma$, $3\sigma$ and $5\sigma$ significance for gaussian
statistics), as estimated by running $10^5$ Monte Carlo simulations.}
\label{fig:2pbl}	
\end{figure}

In Figure~\ref{fig:skybl} we plot the positions on the sky
(Hammer-Aitoff projections in equatorial coordinates) of both the
UHECRS and the selected objects in order to give a visual impression
of how the coincidences arise, particularly for the `doublets' and
`triplet' in the AGASA data. The two-point correlation function for
the four cases are shown in Figure~\ref{fig:2pbl}, calculated according
to the Tinyakov \& Tkachev prescription
\cite{Tinyakov:2001ic,Tinyakov:2001nr,Tinyakov:2001ir}, adopting the
same angular bin size of $2.5^\circ$. To determine the significance of
these correlations we run $10^5$ Monte Carlo simulations, as they did,
to calculate the probability of chance coincidences. It is evident
that while there is indeed a $\sim3\sigma$ correlation between UHECRs
and BL Lacs if suitable cuts are employed \cite{Tinyakov:2001nr}, the
significance weakens to $\sim2.7\sigma$ if the energy cut is relaxed,
and {\em disappears} if the cuts on BL Lacs are also relaxed. Thus
there is {\em no} basis for the claim that BL Lacs are the sources of
UHECRs; indeed cosmologically distant GRBs correlate just as well with
post-GZK UHECRs as do BL Lacs!

We now give details of the four cases considered, with reference to
Table~\ref{table:bllacs}.

\begin{table}[ht]
\begin{center}
Model 1
\end{center}
\begin{tabular}{|c|c|c|c|c|} \hline
& Experimental & Monte Carlo & $95\%$ c.l. upper & ${\cal P}
(>{\rm Exp}$) \\ 
& Value & Mean & Bound & \null \\ \hline\hline 
Clusters & 5 & 0.78335 &  2 & 0.00152 \\ \hline 
Singlets & 1 & 0.75253 &  2 & 0.53514 \\ \hline 
Doublets & 2 & 0.01522 &  0 & 0.00019 \\ \hline 
Triplets & 0 & 0.00014 &  0 & 1       \\ \hline
\end{tabular}
\begin{center}
Model 2
\end{center}
\begin{tabular}{|c|c|c|c|c|} \hline
& Experimental & Monte Carlo & $95\%$ c.l. upper & ${\cal P}
(>{\rm Exp}$) \\ 
& Value & Mean & Bound & \null \\ \hline\hline 
Clusters & 5 & 1.16664 & 3 & 0.00713 \\ \hline 
Singlets & 1 & 1.09889 & 3 & 0.67605 \\ \hline 
Doublets & 2 & 0.03315 & 0 & 0.00061 \\ \hline 
Triplets & 0 & 0.00047 & 0 & 1      \\ \hline
\end{tabular}
\begin{center}
Model 3
\end{center}
\begin{tabular}{|c|c|c|c|c|} \hline
& Experimental & Monte Carlo & $95\%$ c.l. upper & ${\cal P}
(>{\rm Exp}$) \\ 
& Value & Mean & Bound & \null \\ \hline\hline 
Clusters & 9 & 9.50677 & 15 & 0.63391 \\ \hline 
Singlets & 5 & 8.96987 & 14 & 0.94949 \\ \hline 
Doublets & 2 & 0.26078 &  1 & 0.03164 \\ \hline 
Triplets & 0 & 0.00499 &  0 & 1 \\ \hline
\end{tabular}
\begin{center}
Model 4
\end{center}
\begin{tabular}{|c|c|c|c|c|} \hline
& Experimental & Monte Carlo & $95\%$ c.l. upper & ${\cal P}
(>{\rm Exp}$) \\ 
& Value & Mean & Bound & \null \\ \hline\hline 
Clusters & 44 & 46.3406 & 58 & 0.65727 \\ \hline 
Singlets & 36 & 43.8211 & 55 & 0.90202 \\ \hline 
Doublets &  4 & 1.22408 &  4 & 0.05606 \\ \hline 
Triplets &  0 & 0.02351 &  0 & 1 \\ \hline
\end{tabular}
\caption{The observed number of coincidences (within $2.5^\circ$)
between BL Lacs (Models 1, 2 and 3) or GRBs (Model 4) and UHECRs
detected by AGASA is tabulated in total, as well as separately as
coincidenes with single events, doublets and triplets of UHECRs. These
are compared with the results of $10^5$ Monte Carlo simulations
assuming random UHECR arrival directions. The table gives the mean
number of coincidences in the simulations, as well as the 95\%
c.l. upper bound, and the probability $P(>{\rm Exp})$ of attaining
results at least as correlated as the actual data}.
\label{table:bllacs}
\end{table}

{\bf Model 1:} 

%This corresponds to the Tinyakov and Tkachev analysis. There are 39
%events with $E>4.8\times10^{19}$~eV in the AGASA catalogue
%\cite{Hayashida:2000zr}
Of the 22 BL Lacs with $z>0.1$, $m<18$ and $F_6>0.17$~Jy considered by
Tinyakov and Tkachev \cite{Tinyakov:2001nr}, 20 are visible to AGASA
\cite{Hayashida:2000zr} which reported 39 events with
$E>4.8\times10^{19}$~eV. Adopting an angular width of $2.5^\circ$ for
each bin (corresponding approximately to the experimental angular
resolution) only 0.8 coincidences are expected on average while 5 are
observed, the chance probability for which is 0.15\%. It is the
coincidences of 2 BL Lacs (RX J10586+5628 and 2EG J0432+2910) with
UHECR doublets which contributes most of this signal, which has a
significance of $\sim3\sigma$ if the statistics are gaussian. No
coincidences with triplets are seen either in the data or in the Monte
Carlo; note that the triplet coincidence emphasized by Tinyakov and
Tkachev \cite{Tinyakov:2001nr} was composed of Yakutsk events well
below the GZK energy, having arrival directions uncertain by
$>4^\circ$ \cite{yakutsk}.

{\bf Model 2:} Retaining the cuts in the BL Lacs, we now consider all
58 events with $E>4\times10^{19}$eV in the AGASA catalogue
\cite{Hayashida:2000zr}. The probability of clustering has now
decreased by a factor of $\sim5$ and has a significance of only
$\sim2.7\sigma$, being still mostly due to the coincidences of the 2
BL Lacs with UHECR doublets.

{\bf Model 3:} Next we consider all 172 BL lacs \cite{veron2001} which
are visible to AGASA. The correlation has now disappeared completely!
The 2 coincidences of BL Lacs with UHECR doublets now has an accidental
probability of 6.3\%.

{\bf Model 4:} Finally we consider the correlation with 915 GRBs in
the BATSE catalogue \cite{grb} which are visible to AGASA. The (lack
of) correlation is just as significant as for the full set of BL
Lacs. There are 4 coincidences of GRBs (4B~920617C, 4B~931211,
4B~950131 and 4B~960128) with 3 UHECR doublets (having energies
$(21.3, 5.07)\times10^{19}$~eV, $(5.47, 4.89)\times10^{19}$~eV, and
$(5.50, 7.76)\times10^{19}$~eV) --- the first and third GRBs coincide
with the same doublet, while the last GRB coincides with two of the
three events forming the AGASA triplet.

\begin{figure}[hbt]
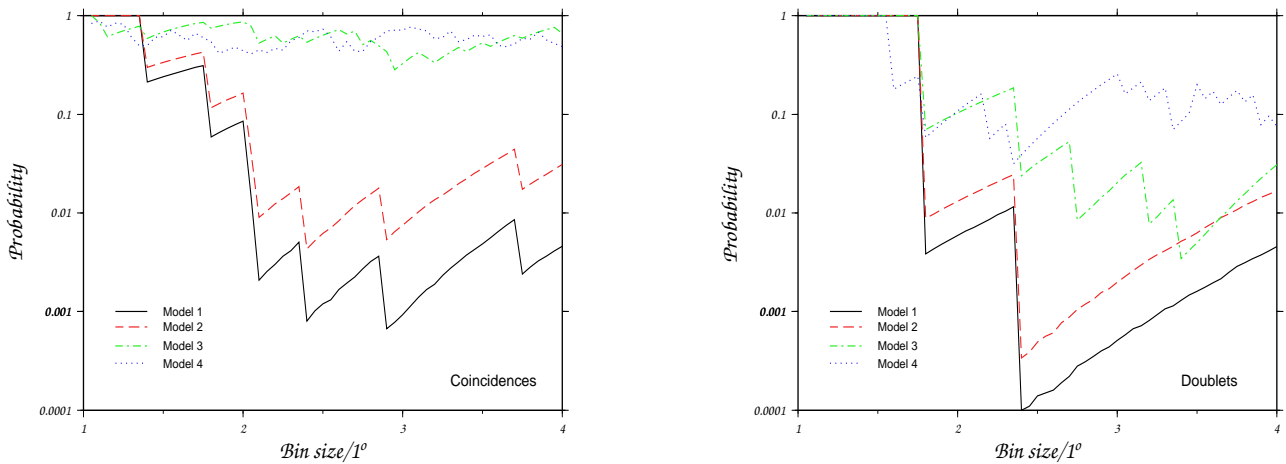

\begin{center}
\begin{tabular}{c@{\hspace{2cm}}c}
\includegraphics[width=7.5cm,height=6.5cm]{fig3a.ps}&
\includegraphics[width=7.5cm,height=6.5cm]{fig3b.ps}
\end{tabular}
\end{center}
\caption{Dependence on the angular bin width of the probability for
coincidences with all UHECRs (left), and doublets of UHECRs (right),
for Model 1 (solid line), Model 2 (long-dashed line), Model 3
(short-dashed line), and Model 4 (dotted line).}
\label{fig:angbl}	
\end{figure}

In Figure~\ref{fig:angbl} we show the dependence of the correlation
probability on the angular bin width for all 4 models. There is indeed
a minimum at $2.5^\circ$ corresponding to the AGASA angular resolution
if the cuts employed by Tinyakov and Tkachev \cite{Tinyakov:2001nr}
are imposed (Model 1). However we see that the significance decreases
as we relax the cut on AGASA events (Model 2) and disappears
altogether if we remove the cuts on the BL Lacs (Model 3). There is
similarly no minimum for correlations with GRBs (Model 4). We conclude
that Tinyakov and Tkachev \cite{Tinyakov:2001nr} vastly underestimated
the `penalty factor' corresponding to the arbitrary cuts they imposed
{\em post facto} on the data.

\section{Clustering and halo dark matter}

\begin{figure}[hbt]
\begin{center}
\begin{tabular}{c@{\hspace{2cm}}c}
\includegraphics[width=7.5cm,height=4cm]{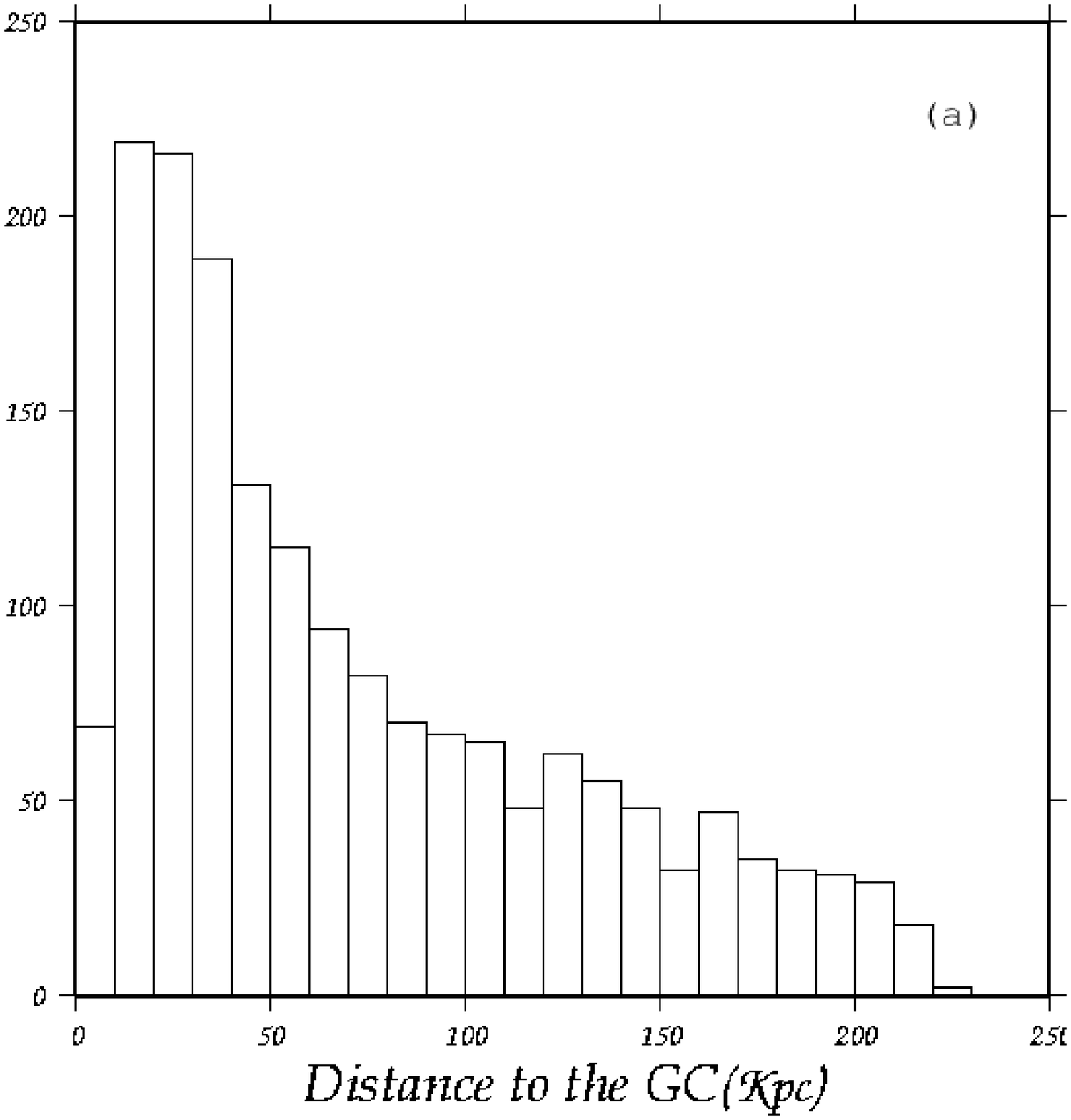}&
\includegraphics[width=7.5cm,height=4cm]{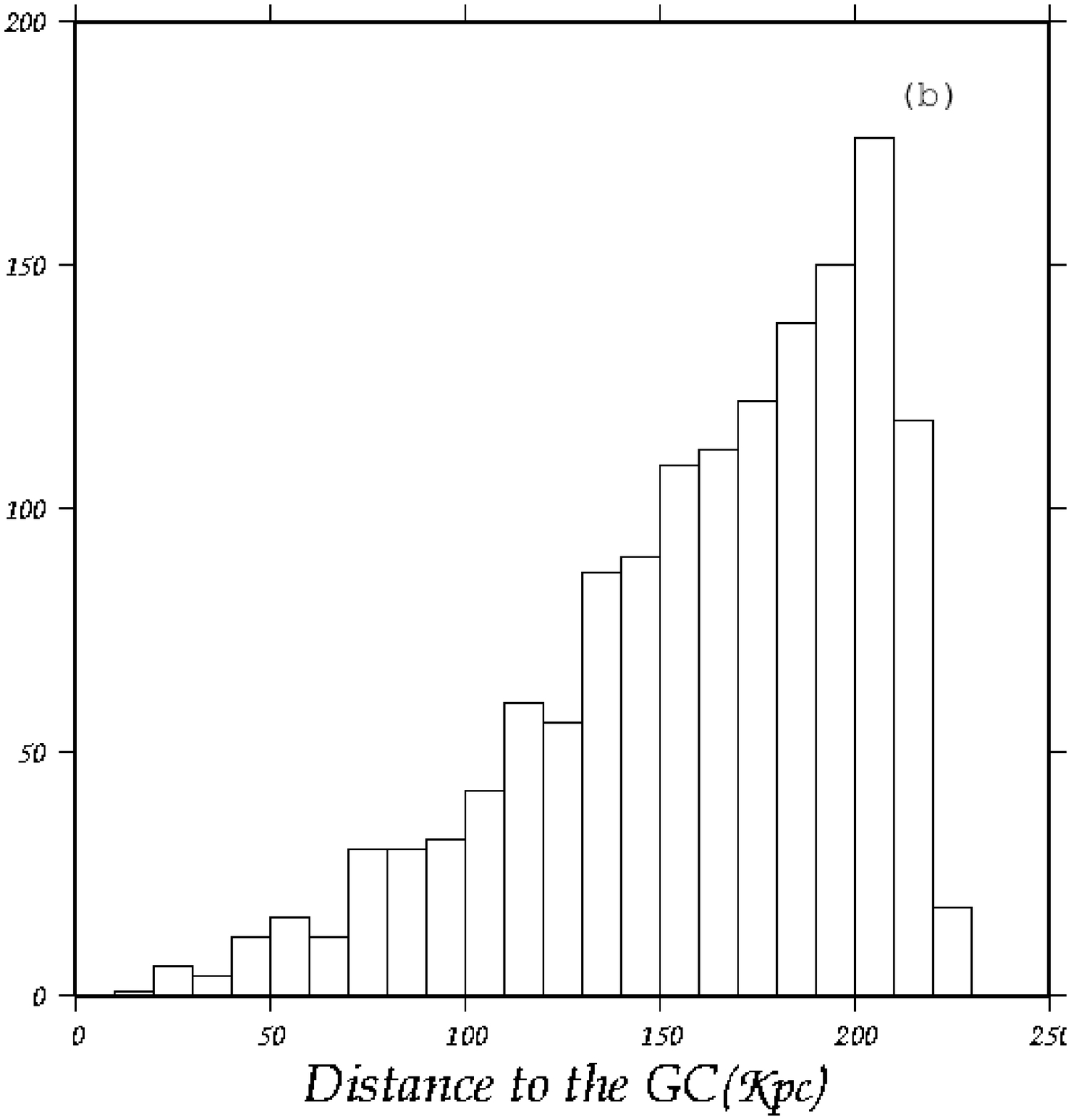}\\
\end{tabular}
\end{center}
\caption{Distribution of the clumps with radius adopted (i) by Blasi
\& Sheth \protect\cite{Blasi:2000ud,Blasi:2001kr}, and (ii) in this
work.}
\label{fig:histgc}	
\end{figure}
\begin{figure}[hbt]
\begin{center}
\begin{tabular}{c@{\hspace{2cm}}c}
\includegraphics[width=7.5cm,height=4cm]{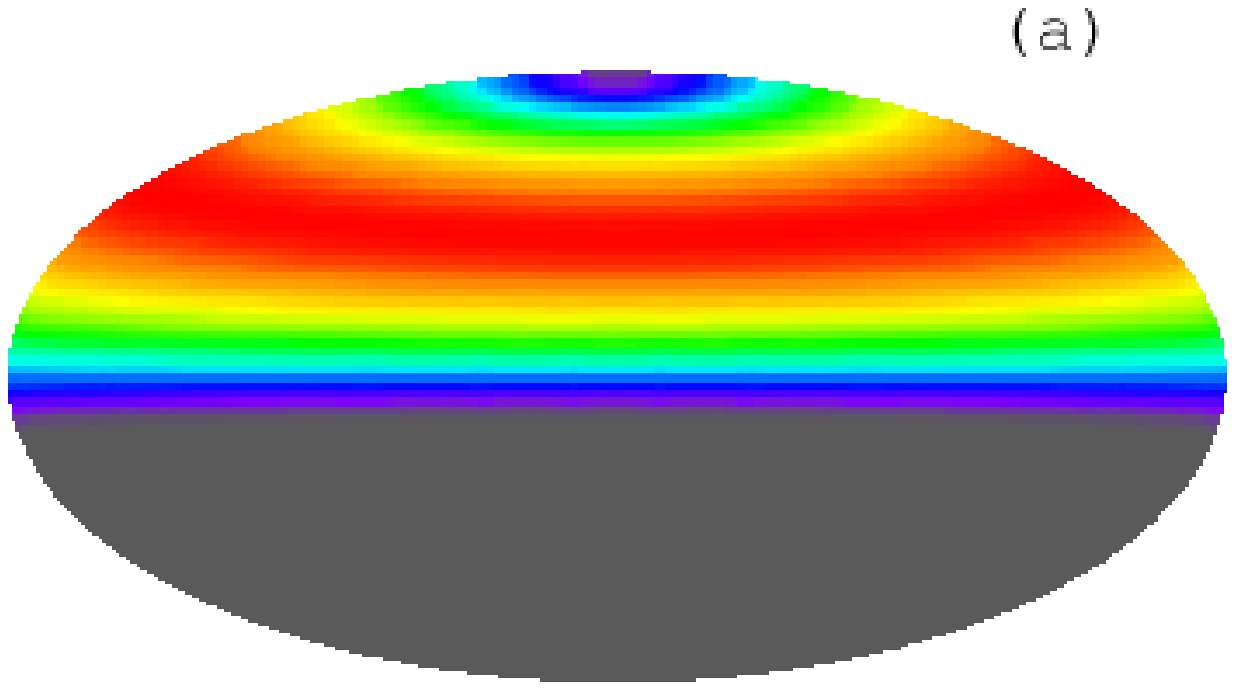}&
\includegraphics[width=7.5cm,height=4cm]{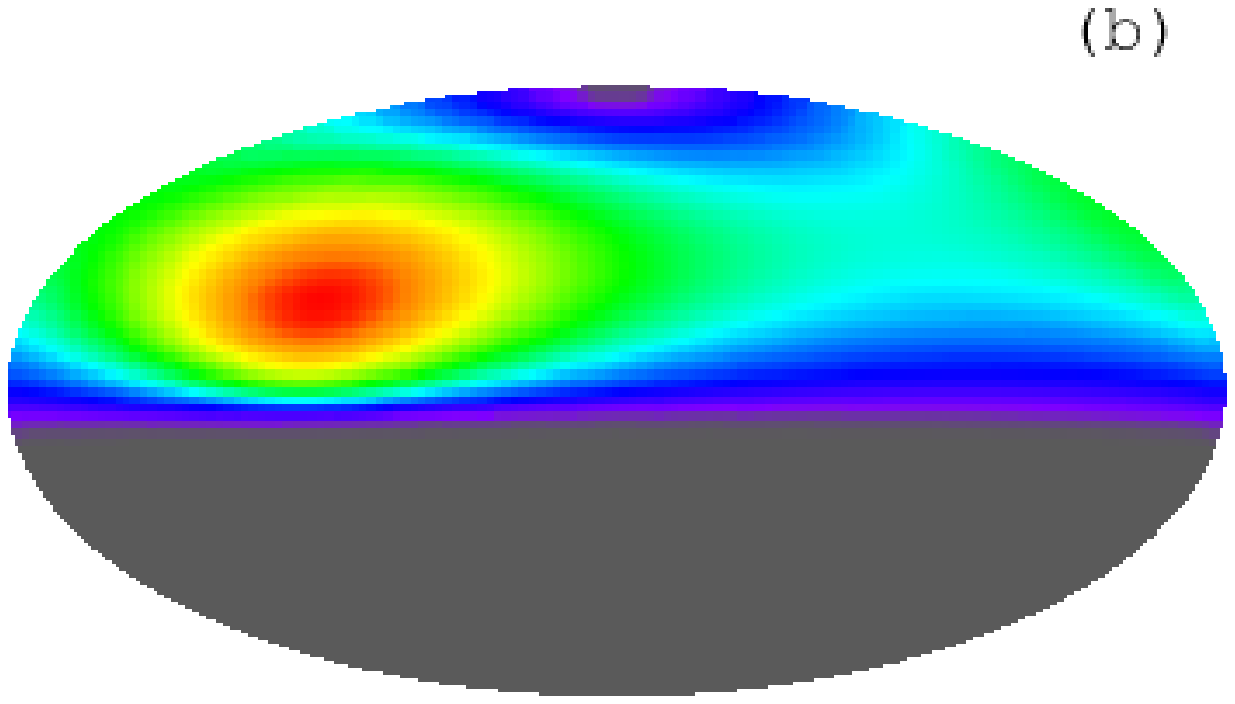}\\
\includegraphics[width=7.5cm,height=4cm]{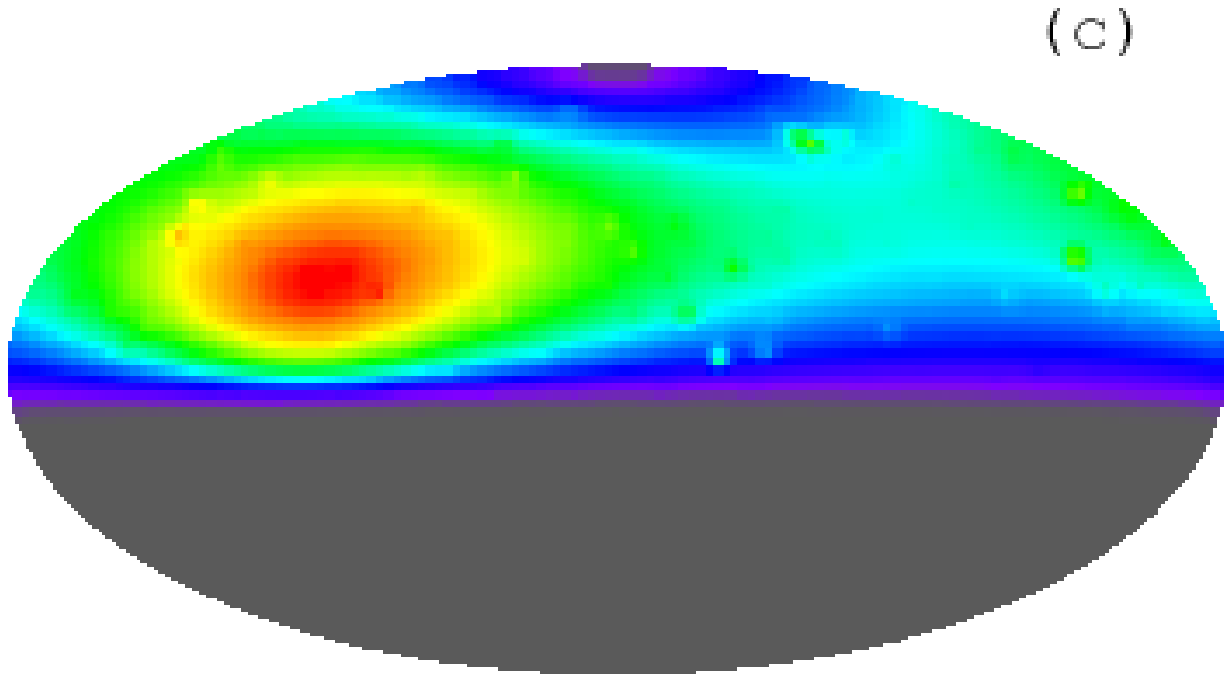}&
\includegraphics[width=7.5cm,height=4cm]{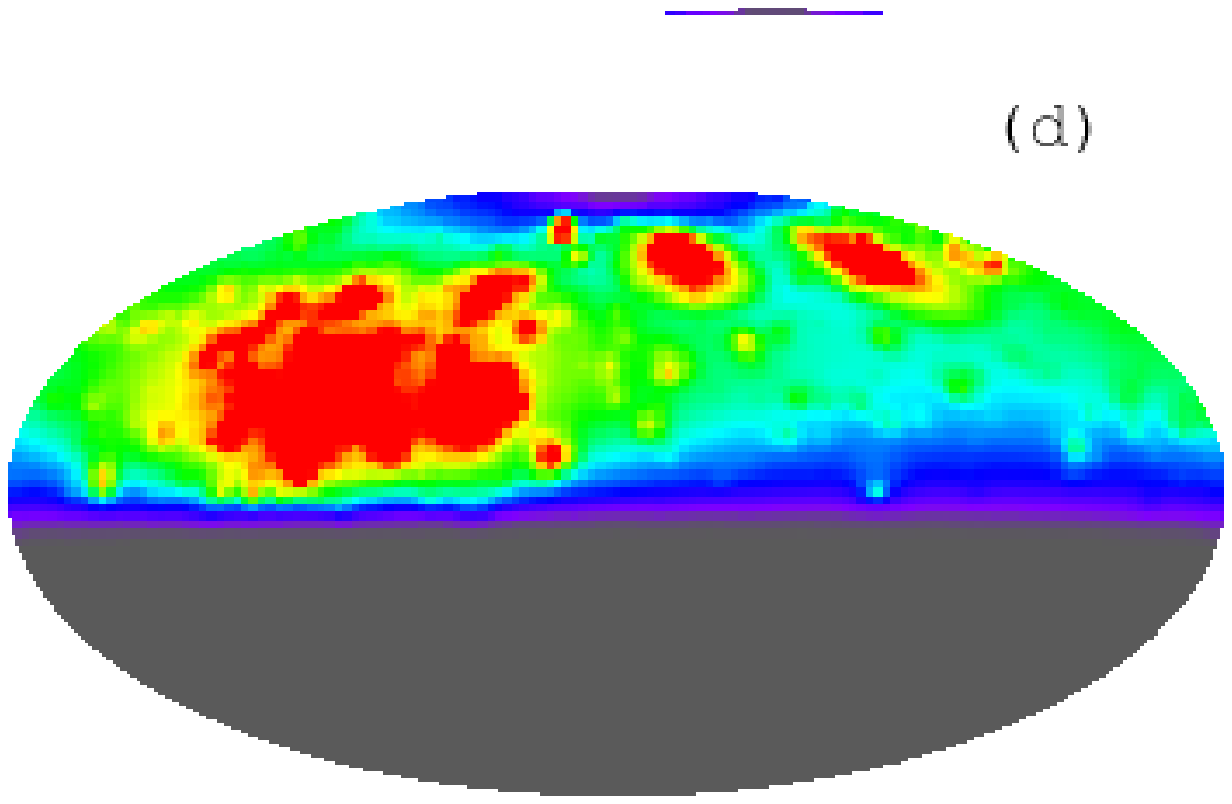}
\end{tabular}
\end{center}
\caption{The UHECR sky as it would be seen by AGASA --- top left:
Isotropic model; top right: Smooth NFW Galactic halo; bottom left:
Smooth NFW Galactic halo with 10\% of the mass in NFW clumps; bottom
right: Smooth NFW Galactic halo with 10\% of the mass in SIS
clumps. The highest flux in each panel corresponds to the darkest (red)
regions, the lowest flux to the lightest (lilac).}
\label{fig:skyag}
\end{figure}

\begin{table}[ht]
\begin{center}
(a) Isotropic arrival directions
\end{center}
\begin{tabular}{|c|c|c|c|c|} \hline
& Experimental & Monte Carlo & $95\%$ c.l. upper & ${\cal P}
(>{\rm Exp}$) \\ 
& Value & Mean & Bound & \null \\ \hline 
Clusters & 8 & 1.60217 & 4 & 0.00038 \\ \hline 
Doublets & 8 & 1.59989 & 4 & 0.00029 \\ \hline 
Triplets & 1 & 0.01661 & 0 & 0.01642 \\ \hline
\end{tabular}
\begin{center}
(b) Galaxy halo modelled by a smooth NFW profile
\end{center}
\begin{tabular}{|c|c|c|c|c|}\hline
& Experimental & Monte Carlo & $95\%$ c.l. upper & ${\cal P}
(>{\rm Exp}$) \\ 
& Value & Mean & Bound & \null \\ \hline 
Clusters & 8 & 1.70778 & 4 &  0.00094 \\ \hline 
Doublets & 8 & 1.71073 & 4 & 0.00076 \\ \hline 
Triplets & 1 & 0.02003 & 0 & 0.01985 \\ \hline
\end{tabular}
\begin{center}
(c) Galaxy halo modelled by a smooth NFW profile plus NFW profile clumps
\end{center}
\begin{tabular}{|c|c|c|c|c|}\hline
& Experimental & Monte Carlo & $95\%$ c.l. upper & ${\cal P}
(>{\rm Exp}$) \\ 
& Value & Mean & Bound & \null \\ \hline 
Clusters & 8 & 1.70712 & 4 & 0.00083 \\ \hline 
Doublets & 8 & 1.70282 & 4 & 0.00057 \\ \hline 
Triplets & 1 & 0.02023 & 4 & 0.02001 \\ \hline
\end{tabular}
\begin{center}
(d) Galaxy halo modelled by a smooth NFW profile plus SIS clumps
\end{center}
\begin{tabular}{|c|c|c|c|c|}\hline
& Experimental & Monte Carlo & $95\%$ c.l. upper & ${\cal P}
(>{\rm Exp}$) \\ 
& Value & Mean & Bound & \null \\ \hline 
Clusters & 8 & 2.1306 & 5 & 0.056 \\ \hline 
Doublets & 8 & 2.2589 & 5 & 0.029 \\ \hline 
Triplets & 1 & 0.057  & 1 & 0.055 \\ \hline
\end{tabular}
\caption{The experimentally observed numbers of clusters, doublets and
triplets are compared with the results of $10^5$ Monte Carlo
simulations. The table gives the mean numbers of clusters,
doublets and triplets for the simulations, the 95 \% upper bounds,
together with the probability $P(>{\rm Exp})$ of obtaining results at
least as clustered as the data }.
\label{table:probs}
\end{table}
\begin{figure}[hbt]
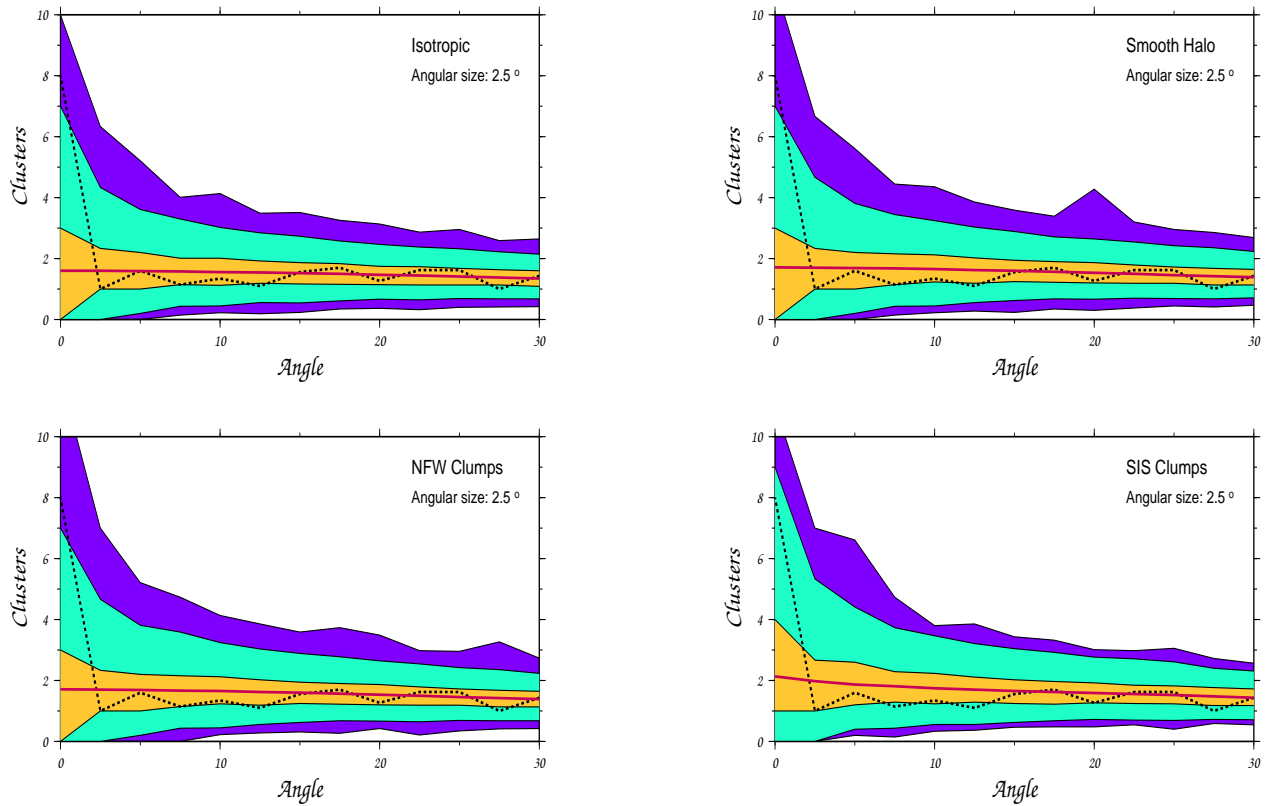

\begin{center}
\begin{tabular}{c@{\hspace{2cm}}c}
\includegraphics[width=7.5cm,height=5.5cm]{fig6a.ps}&
\includegraphics[width=7.5cm,height=5.5cm]{fig6b.ps}\\
\includegraphics[width=7.5cm,height=5.5cm]{fig6c.ps}&
\includegraphics[width=7.5cm,height=5.5cm]{fig6d.ps}
\end{tabular}
\end{center}
\caption{The two point correlation function for UHECRs generated with
(a) isotropic arrival directions, (b) in a smooth NFW halo model, (c)
in a smooth NFW halo with 10 \% of the mass in NFW clumps or (d) in
SIS clumps. The angular size on which correlations are sought is
$\Delta\theta=2.5^\circ$. The thick dashed line corresponds to the
autocorrelation function of the experimental data. The thick unbroken
line is the average autocorrelation function of $10^5$ Monte Carlo
simulations. The regions shaded light, medium and dark correspond to
chance probabilities less than 31.73\%, 0.27\% and 0.0057\% (chosen so
as to correspond to $1\sigma$, $3\sigma$ and $5\sigma$ significance
for gaussian statistics), as estimated by running $10^5$ Monte Carlo
simulations.}
\label{fig:tp25}	
\end{figure}

In cold dark matter cosmogonies, galaxies are built up from the
merging and accretion of smaller structures. This leaves phase space
substructure in the form of clumps and streams imprinted in galaxy
haloes today. Numerical simulations of galaxy formation suggest that
$\lesssim10\%$ of the total halo mass may be in the form of such
substructure \cite{Moore:1999wf,Ghigna:1999sn}. However, care is
needed in interpreting the results of such simulations, as very high
resolution is required to resolve the substructure left over from
merger events.  More directly, there is unambiguous evidence of
substructure in the stellar populations of both the Galactic
\cite{ArnoldGilmore:1990,Ibata:1994} and the Andromeda (M31) haloes
\cite{Ibata:2001}. These are probably the remnants of smaller galaxies
engulfed by larger neighbours. The anomalous flux ratios of quadruplet
lenses have also been claimed as evidence of
substructure \cite{Kochanek:2001dx,Metcalf:2001es}, but this evidence
is not clear-cut, as microlensing, differential extinction or
scatter-broadening may also be affecting the flux ratios
\cite{EvansWitt:2002}.

In the inner parts of the Galactic halo (say $r\lesssim25$~kpc),
dynamical friction and tides are efficient at erasing substructure
\cite{Ghigna:1999sn}. In the outer parts, clumps and streams can
preserve their identity for longer. Blasi \& Sheth
\cite{Blasi:2000ud,Blasi:2001kr} have suggested that the clumping of
halo dark matter will give rise to clustering on the UHECR sky if the
UHECRs themselves arise from the decay of superheavy dark matter
particles \cite{Berezinsky:1997hy,Birkel:1998nx}. Blasi \& Sheth
developed models in which the clumps occur with masses $m$ at a
distance $r$ from the center according to the joint probability
distribution
\begin{equation}
n_{\rm cl} \propto \left( {1 \over m^{1.9}} \right) \left( {1 \over 1
+ (r/r_{\rm c})^2} \right)^{3/2},
\end{equation}
where $r_{\rm c}$ is a constant, which they take as $\sim10$ kpc. They
argue that this is a good fit to the simulation data. Nonetheless, it
runs counter to physical intuition, as the number density of clumps
peaks in the center rather than the outlying portions of the Galaxy
halo. Blasi \& Sheth assume that the clumps themselves have the
singular isothermal sphere (SIS) density law $\rho_{\rm
cl}\propto\,r_{\rm cl}^{-2}$, where $r_{\rm cl}$ is the radial
coordinate measured from the clump center. The singular isothermal
spheres are truncated at their tidal radius in the Galaxy halo, which
is assumed to take the Navarro-Frenk-White (NFW) form
\cite{Navarro:1996}
\begin{equation}
\rho_{\rm NFW} \propto { 1 \over r ( 1 + r/r_{\rm s} )^2 }.
\end{equation}
The choice of the isothermal law for the clumps has no physical basis
whatsoever. In fact, the results of numerical simulations are usually
claimed to be {\em self-similar}, in the sense that superclusters,
clusters, galaxy haloes and sub-haloes all have NFW
profiles \cite{Navarro:1996}.

To test the robustness of Blasi \& Sheth's results, we develop a
different model of the substructure. First, the clumps are distributed
homogeneously in the Galaxy halo, so the number of clumps within a
radius $r$ increases like $r^3$. Second, the masses of the clumps are
chosen so that there is equal mass in equal decades (i.e.,
$n(m)\propto\,m^{-2}$).  Third, the clumps themselves are chosen to
have NFW profiles in the parent NFW halo of the Galaxy. This is
motivated by the scale-freeness of the results of the simulations.  In
fact, in the inner parts of the Milky Way, there is ample evidence
that the halo does {\em not} have the NFW form
\cite{Evans:2000}. However, our main motivation here is to understand
the results obtained by Blasi \& Sheth, so we have only changed the
substructure properties and not the underlying smooth halo model. (For
reference, the NFW concentration parameter $c$ is $\sim10$ for the
Galaxy halo and $\sim5$ for the clumps.) From the simulations, it is
found that most clumps fall in at a redshift of $z\sim4$ and their
concentration remains largely frozen after this. The overdensity at
formation is calibrated with respect to the critical density at that
epoch.  In selecting parameters for the clump distribution, our
canonical choice is to distribute a generous 10 per cent of the total
halo mass in clumps between $10^7$ and $10^{10}$~M$_\odot$. Given the
mass of the clump, the virial radius follows. The virial radius and
the concentration determine the NFW lengthscale $r_{\rm s}$.

\begin{figure}[t]
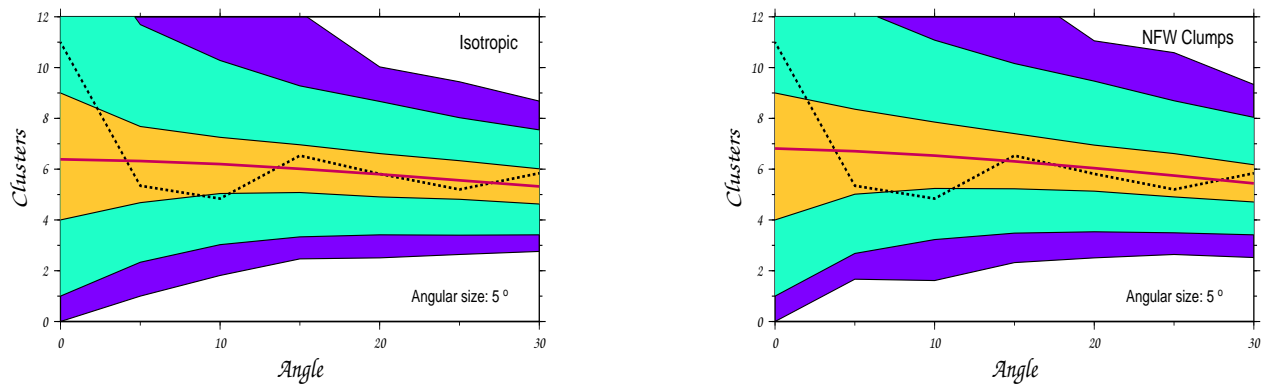

\begin{center}
\begin{tabular}{c@{\hspace{2cm}}c}
\includegraphics[width=7.5cm,height=5.5cm]{fig7a.ps}&
\includegraphics[width=7.5cm,height=5.5cm]{fig7b.ps}
\end{tabular}
\end{center}
\caption{As Figure~\ref{fig:tp25}, but changing the angular size
$\Delta \theta$ to $5^\circ$ Only two models are shown, namely
isotropic arrival directions (left) and a smooth NFW model with NFW
clumps (right).}
\label{fig:tp5}	
\end{figure}
\begin{figure}[hbt]
\begin{center}
\includegraphics[width=7.5cm,height=6.5cm]{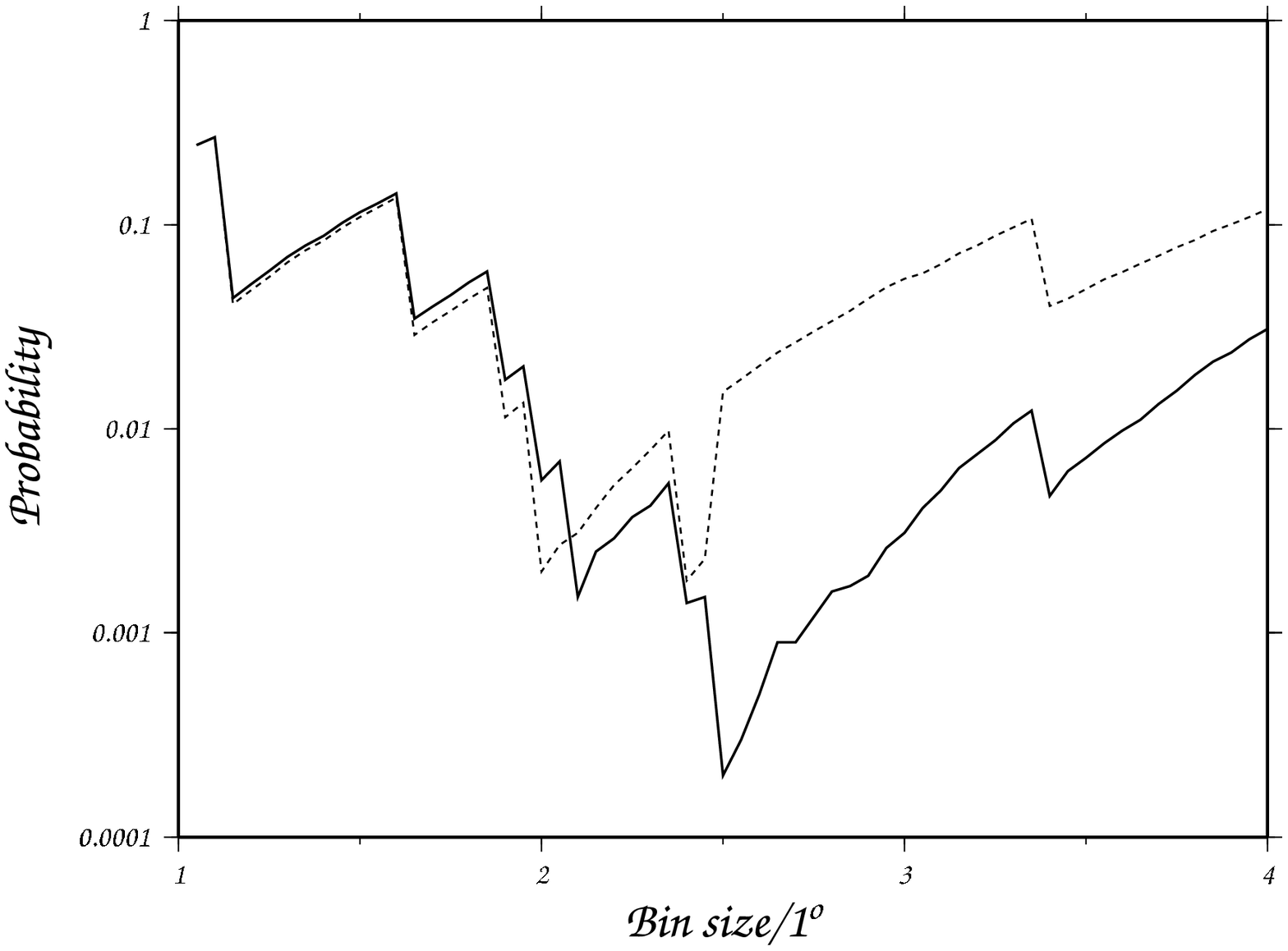}
\end{center}
\caption{The probability of obtaining as many clusters (solid line)
and doublets (dashed line) as the data given as a function of the
angular size $\Delta\theta$ of the correlations.}
\label{fig:bina}	
\end{figure}

Let us start by presenting different properties of the clump
populations generated according to the recipes of Blasi \& Sheth and
ourselves. For a Galaxy halo, we have typically the same numbers of
clumps ($\sim1600$) in both models. With our prescription, the clumps
tend to be smaller in extent which makes their detection more
difficult. However, the main difference is in the distribution of
clumps with Galactocentric radius, as shown in
Figure~\ref{fig:histgc}.  In Blasi \& Sheth's model, the peak in the
radial distribution of the clumps occurs at $\sim20$~kpc, which is
close to the Solar circle.  This is of course optimum for causing
visible consequences. In our model, the peak occurs at 220 kpc, very
much in the outer parts of the Galaxy halo. Let use remark that the
numerical simulations clearly show that most of the surviving
substructure {\em is} in the outer parts. Both NFW and SIS profiles
are singular at the origin, so a small regularizing core radius must
be included in the computations. For the NFW profiles, we take our cue
from our earlier paper \cite{Evans:2001rv} where the regularizing core
$r_\epsilon$ was chosen as $\sim 0.5$ kpc (the resolution limit) for a
halo of $\sim250$~kpc extent. For the NFW clumps, $r_\epsilon$ is
scaled to be the same fraction of the extent. For the SIS profiles,
we adopt $r_\epsilon=3\times10^{-7}$~kpc, as suggested by the limit
imposed by particle dark matter self-annihilation
\cite{Blasi:2002ct,Tyler:2002ux}.

Figure~\ref{fig:skyag} shows the incoming UHECR flux in a Hamer-Aitoff
projection in equatorial coordinates folded with the response of the
AGASA detector. The flux from the underlying smooth model is
calculated by integrating the emissivity density along the line of
sight (see e.g., Ref.~\cite{Evans:2001rv}). However, the angular size
of clumps can be smaller than the angular resolution of the
instrument, in which case the flux is computed by the volume integral
of the emissivity over the clump, divided by the square of the
distance of the clump from Earth.  There are four panels showing
random arrival directions, a smooth NFW galactic halo, a smooth NFW
halo plus NFW clumps and a smooth halo plus SIS clumps. The grey
region corresponds to no detection, as AGASA is a Northern Hemisphere
experiment.  The lower right panel shows irregularities,
caused by the SIS clumps which are brighter than any contribution from
the underlying smooth halo. The lower left panel corresponding to NFW
clumps shows much less evidence for irregularities. The flux is much
more uniform with less clustering.  We cannot verify the claim made by
Blasi \& Sheth that a smooth NFW halo alone is able to provide almost
half the observed clustering \cite{Blasi:2000ud,Blasi:2001kr}. In
fact, it is hard to see how a smooth halo can be responsible for small
scale flux variations.

To quantify this, we use the 58 events in the AGASA experiment above
$4\times10^{19}$eV as our dataset. The two point autocorrelation
function for the four cases is shown in Figure~\ref{fig:tp25}. Samples
of 58 UHECRs, with the AGASA response function folded in, are
generated, and the average autocorrelation is compared to the one
found for the experimental data.  The clustering in the experimental
dataset is not well reproduced. Even when SIS clumps are present, the
disagreement is at the $2.8 \sigma$ level and it is beyond $3\sigma$
for the other models. We also note from panels (a), (b) and (c) that
the smooth NFW profile (with or without clumps) is almost
indistinguishable from the isotropic model. Table~\ref{table:probs}
gives the explicit probabilities deduced from $10^5$ Monte Carlo
calculations.  We see that in the first three models, we fall short of
obtaining the 8 clusters required as less than 2 are expected on
average.  The probability of obtaining 8 or more clumps is of order
$10^{-3}-10^{-4}$. When SIS clumps are introduced in the halo,
however, the probability increases to 0.056. Note that, in every case,
the discrepancy between model predictions and experiment is always
less than $5\sigma$.

The angular width of each bin was taken to be $2.5^\circ$ in the
calculations reported above. If this is enlarged to $5^\circ$, then
the discrepancies between model predictions and data are smaller.
This is a reasonable angular size on which to look for correlations as
it corresponds to the typical deflection of a $4\times10^{19}$~eV
proton in the Galactic magnetic field
\cite{Stanev:1997,MedinaTanco:1997rt}. Figure~\ref{fig:tp5} shows the
autocorrelation functions for the cases of isotropic arrival
directions and for a smooth NFW halo with NFW clumps. We expect
typically 6 and 7 clusters for these two cases. The probability of
obtaining as many clusters as the experimental data (11) rises to 0.06
and 0.07 respectively. Note that, even for the isotropic model, the
discrepancy is now at only the $\approx2\sigma$ level.  The dependence
of the probability on the bin width is shown in Figure~\ref{fig:bina}
for the model with NFW clumps in an NFW halo. It reaches a minimum
around $2.5^\circ$. This is where the discrepancy with the data is at
a maximum. Either decreasing or increasing $\Delta\theta$ gives model
predictions in better agreement with the data. So, this Figure shows
how sensitive the clustering results are to the bin width. For
example, changing $2.5^\circ$ to $3^\circ$ causes almost an order of
magnitude change in the probability.

Using the 92 events from the combined datasets of AGASA, Volcano
Ranch, Haverah Park and Yakutsk \cite{Uchihori:1999gu}, Blasi \& Sheth
estimated that the probability of attaining more doublets than the
data from SIS clumps is 12\% (for $3^\circ$ bin widths) and and 47\%
(for $4^\circ$). These high numbers appear to be primarily a
consequence of placing the SIS clumps nearby. Using our model for the
mass and spatial distribution of the clumps, we obtain corresponding
probabilities of 7\% (for $3^\circ$) and 29\% (for $4^\circ$). If the
clumps have an NFW profile as is more likely, these numbers are
reduced further to 3.5\% and 15\% respectively.

We conclude that clustering is {\em not} a generic prediction of the
decaying dark matter model. Even though clumps are indeed expected in
the dark matter distributions in Galaxy haloes, any clustering of
UHECRs depends sensitively on the density profile of the clumps. In
particular, NFW clumps do not give rise to much clustering, but SIS
clumps may do so. However, SIS profiles are not very natural, and
almost all the signal comes from the $r_{\rm cl}^{-2}$ singularity. In
fact, the self-similarity of the structure formation process suggests
that NFW clumps are more natural. We note that the 10\% of mass that
we placed in clumps is already generous, and so it is not clear
whether any real clustering of UHECRs can be expected from dark matter
substructure. However, it is also not clear that there is any real
clustering of the UHECRs at all, as the signal is less than $2\sigma$
at the most natural angular scale of $\Delta\theta=5^\circ$.

\section{Discussion}

It has been a general expectation that it should be possible to
identify the long-sought sources of cosmic rays at energies exceeding
$\sim10^{19}$~eV, when their arrival directions can no longer be
randomised by Galactic magnetic fields. However the sky distribution
has remained consistent with isotropy up to the highest energies
observed. It is clear that at such energies the sources cannot be in
the disk of the Galaxy and all experiments indicate that the energy
spectrum of such sources is significantly flatter than the component
at lower energies. However it is not clear whether the isotropy of
arrival directions implicates a relatively {\em local} population of
sources in the Galactic halo (e.g. decaying supermassive dark matter)
or a cosmologically {\em distant} population of astrophysical
accelerators (e.g. active galaxies or $\gamma$-ray bursts). Support
for the latter possibility has come from the new HiRes data which
shows a cutoff in the spectrum beyond $E_{\rm
GZK}\simeq4\times10^{19}$~eV, as has long been expected for
extragalactic sources. However the AGASA collaboration has reaffirmed
that there is no GZK cutoff in their data, which strongly favours the
former possibility.

Given this confusing situation, the indication of small-angle
clustering in the AGASA data has naturally been seized upon as a
possible further clue as to the nature of the sources. It has been
argued both that the clusters coincide with a specific class of
extragalactic objects, viz. BL Lacs
\cite{Tinyakov:2001nr,Tinyakov:2001ir}, and, alternatively, that such
clustering might arise due to the expected clumping of halo dark
matter \cite{Blasi:2000ud,Blasi:2001kr}. The BL Lac hypthesis is in
fact {\em inconsistent} with the absence of the GZK cutoff in the same
AGASA data since many of the identified objects are at very large
distances. Hence it appeared more plausible that the sources are
clumps of dark matter in the Galactic halo (composed in part of
supermassive decaying particles) which would explain the absence of
the GZK cutoff.

We have shown that the correlations claimed between BL Lacs and the
observed clusters of UHECRs are spurious, being entirely due to
selection effects. This is not the first time that such correlations
with a particular class of astrophysical accelerators has been
claimed; the moral is clearly that care must be taken to not become
intrigued by weak accidental correlations and then make arbitrary cuts
on the dataset to emphasise them further. We have also found that the
extent to which dark matter may be clumped in the halo is not
sufficient to generate the observed small-angle clustering, if the
UHECRs indeed arise from decaying dark matter. Here the proponents
have been misled due to the use of an unphysical density profile for
the clumps, as well as a radial distribution in the Galaxy which is
inconsistent with the general expectations for hierarchical structure
formation.

The net result of our investigations is thus rather negative. The
claimed small-angle clustering in the arrival directions of post-GZK
UHECRs does not definitively implicate either extragalactic compact
sources such as BL Lacs, or decaying clumps of dark matter in the
Galactic halo. On the positive side, the forthcoming increase in
statistics from the Pierre Auger Observatory will enable us to
identify the expected signal from dark matter decays if this is indeed
the source of UHECRs. Moreover Auger will also definitively resolve
the current contradiction between the air shower and atmospheric
fluorescence methods for energy measurement, thus clarifying whether
the spectrum does have a GZK cutoff. If so, searches for coincidences
with cosmologically distant candidate sources such as active galaxies
or $\gamma$-ray bursts would be of interest. Again the increased
statistics provided by Auger would enable the significance of such
coincidences to be meaningfully assessed. We will soon know whether
the mystery of UHECRs implicates astrophysical sources or new physics
beyond the Standard Model.

\begin{acknowledgments}
We are grateful to Motohiko Nagano and Alan Watson for discussions and
for providing UHECR data sets, and to Ben Moore for clarifications of
the results obtained in numerical simulations of halo substructure. We
also thank Pascuale Blasi, Ravi Sheth and Igor Tkachev for helpful
correspondance.
\end{acknowledgments}

\vskip 1truecm
\noindent
NOTE ADDED IN PROOF: While this paper was being refereed, a preprint
(astro-ph/0301336) by Tinyakov and Tkachev appeared claiming that our
critique of their work is not justified. They calculate that the `penalty'
for making a-posteriori cuts on the BL Lac and UHECR catalogues reduces
the significance of the coincidences they find by only a factor of 15. 
However as our Monte Carlo simulations demonstrate directly, this is a
vast underestimate and in fact the claimed coincidences are likely to have
arisen just by chance.

\end{document}